\documentclass[sort&compress,preprint,5p,10pt,twocolumn]{elsarticle}

\usepackage{amssymb}
\usepackage{amsmath}
\usepackage{amsthm}
\usepackage{bm}
\usepackage{xcolor}
\usepackage{enumitem}

\journal{Physics Letters A}

\begin{document}

\begin{frontmatter}

\title{Is it possible to determine unambiguously the Berry phase solely from quantum oscillations?}

\author{Bogdan M. Fominykh, Valentin Yu. Irkhin and Vyacheslav V. Marchenkov}

\affiliation{organization={M. N. Mikheev Institute of Metal Physics, Ural Branch of the Russian Academy of Sciences},
            addressline={S. Kovalevskaya str., 18}, 
            city={Ekaterinburg},
            postcode={620108}, 
            country={Russia}}

\begin{abstract}
The Berry phase, a fundamental geometric phase in quantum systems, has become a crucial tool for probing the topological properties of materials. Quantum oscillations, such as Shubnikov-de Haas (SdH) oscillations, are widely used to extract this phase, but its unambiguous determination remains challenging. This work highlights the inherent ambiguities in interpreting the oscillation phase solely from SdH data, primarily due to the influence of the spin factor $R_S$, which depends on the Landé $g$-factor and effective mass. While the Lifshitz-Kosevich (LK) theory provides a framework for analyzing oscillations, the unknown g-factor introduces significant uncertainty. For instance, a zero oscillation phase could arise either from a nontrivial Berry phase or a negative $R_S$. We demonstrate that neglecting $R_S$ in modern studies, especially for topological materials with strong spin-orbit coupling, can lead to doubtful conclusions. Through theoretical analysis and numerical examples, we show how the interplay between the Berry phase and Zeeman effect complicates phase determination. Additionally, we also discuss another underappreciated mechanism - the magnetic field dependence of the Fermi level. Our discussion underscores the need for complementary experimental techniques to resolve these ambiguities and calls for further research to refine the interpretation of quantum oscillations in topological systems.
\end{abstract}

\begin{keyword}
topological materials \sep quantum oscillations \sep Berry phase \sep $g$-factor \sep Zeeman effect.
\end{keyword}

\end{frontmatter}

\section{Introduction}
\label{sec1}

Shubnikov-de Haas (SdH) oscillations serve as a powerful tool for investigating the electronic structure of materials, enabling the determination of carrier effective mass, Fermi surface parameters, and the oscillation phase. The latter is intimately connected to the Berry phase - a fundamental geometric phase that emerges during cyclic evolution of quantum systems \cite{doi:10.1098/rspa.1984.0023,PhysRevLett.82.2147,Zhao31122022,Vedeneev:2017,pancharatnam1956generalized,Yang31122022,annurev:/content/journals/10.1146/annurev-matsci-070218-010023,Weng04052015,KANE20133,doi:10.1142/S0217732305016579}
\begin{equation}
	\phi_{B}=\oint{\bm{A}}d\bm{k}=\int\Omega d^{2}\bm{k}
\end{equation}
where $\bm{A}=i\langle u_{n}(\bm{k})|\nabla_{\bm{k}}|u_{n}(\bm{k})\rangle$ is Berry connection and $\Omega=\nabla_{\bm{k}}\times\bm{A}=i\langle\nabla_{\bm{k}}u_{n}(\bm{k})|\times|\nabla_{\bm{k}}u_{n}(\bm{k})\rangle$ is Berry curvature ($u_{n}(\bm{k})$ is Bloch wave function). For topological materials, the Berry phase equals $\pi$ when an electron traverses a closed loop around a Dirac or Weyl singular point in the Brillouin zone (Fig. 1a). The Berry phase has a natural geometric analogue that emerges during parallel transport of vectors along closed contours on curved surfaces. Here, parallel transport refers to moving a vector along a surface path while keeping it tangent to the local surface plane at all times. Consider a vector transported along a closed loop C = (1$\rightarrow$2$\rightarrow$3$\rightarrow$1) on a sphere (Fig. 1b). If the vector initially points along a specific direction, completing this loop will return it to the starting point with a different orientation. The final vector becomes rotated by an angle $\phi$ relative to its initial position. This geometric phase accumulation mirrors the behavior of the Berry phase in quantum systems.
The discovery of this concept has made it possible to explain many effects in condensed matter physics, such as a topological electric polarization \cite{SPALDIN20122}, weak antilocalization \cite{nano11051077}, chiral anomaly \cite{FUJIKAWA2023103992}, anomalous Hall effect \cite{doi:10.7566/JPSJ.92.124702}, etc.

\begin{figure}
	\centering
	\includegraphics[width=1\columnwidth]{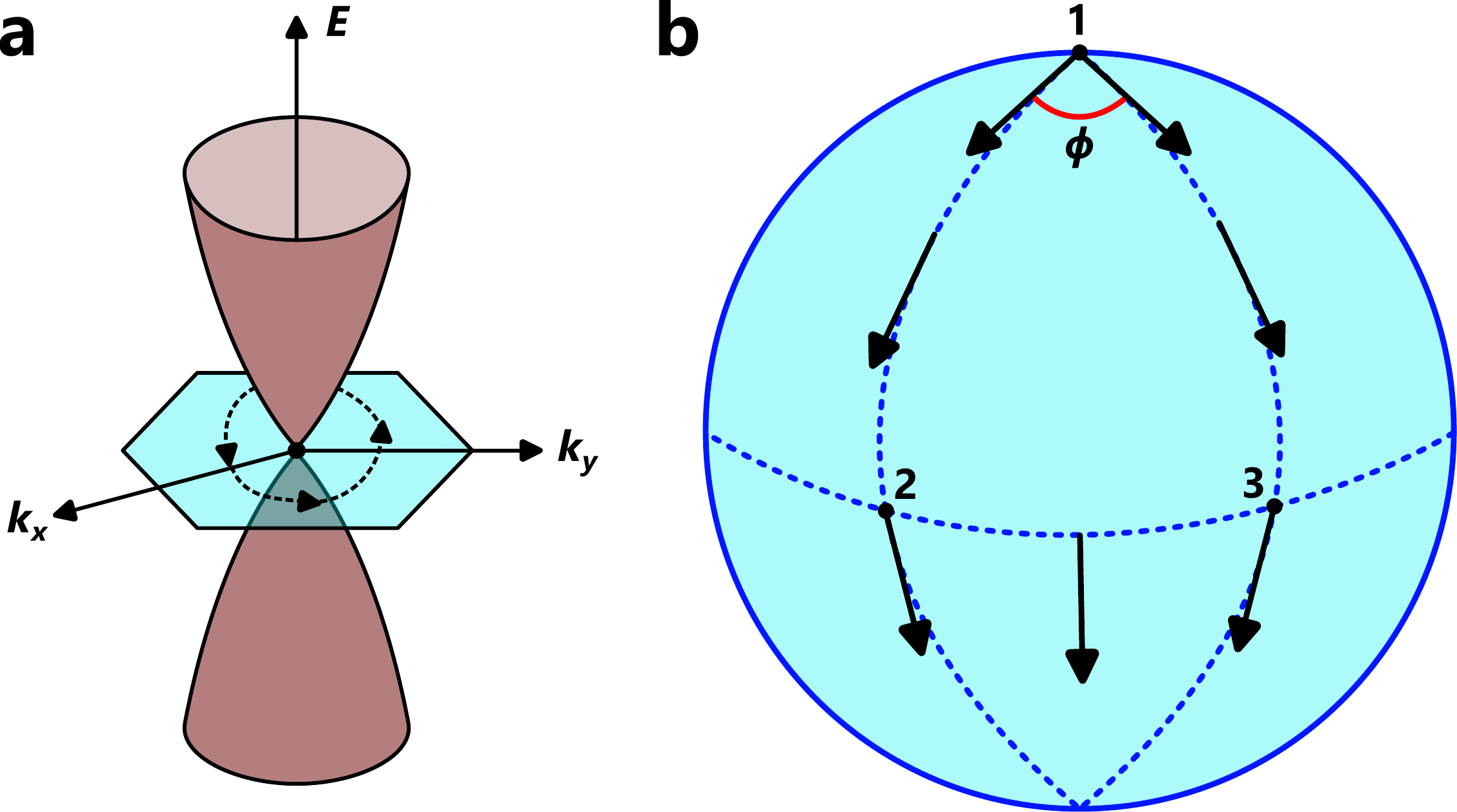}
	\caption{a) Schematic illustration of a possible closed trajectory (dashed line) in the surface Brillouin zone around a Dirac point. Electrons moving along such a path acquire a nontrivial $\pi$ Berry phase. b) Parallel transport of a vector along a closed loop on a sphere. } 
	\label{fig1}
\end{figure}

In contemporary research, the Berry phase has become a subject of comprehensive study as it provides a straightforward means to assess the topological properties of band structures \cite{ren2021local,doi:10.1126/science.1189792,PhysRevLett.113.246402,PhysRevX.5.031023,Ding_2021,doi:10.1073/pnas.1618004114,liu2017magnetic,busch2018high,10.1063/5.0168564,ROMANOVA201843,10.1002/pssr.201510260,BARUA201668,PAN201613,vedeneev2016berry,fominykh2025non,PhysRevB.110.155113,Zhang_2017,Cao_2022,Wang_2017}. However, despite the apparent simplicity of oscillation interpretation, extracting the Berry phase solely from quantum oscillation data presents several fundamental challenges.

According to the Lifshitz-Kosevich theory, the oscillation phase depends not only on the Berry phase $\phi_{B}=2\pi\beta$, but also on the spin factor $R_{S}$, which is determined by the Land\'e $g$-factor and effective mass. Since the $g$-factor typically remains unknown in standard transport measurements, this introduces significant ambiguity in experimental data interpretation. For instance, a zero oscillation phase could indicate either a nontrivial Berry phase ($\beta=0.5$) or a negative $R_{S}$ value resulting from specific $g$-factor characteristics.

Early studies employed the sign of $R_{S}$ to constrain possible $g$-factor values. However, in modern research -- particularly for topological materials with strong spin-orbit coupling -- neglecting this factor may lead to erroneous conclusions. This work is motivated by our previous study \cite{FOMINYKH2025182966}, where we investigated quantum oscillations of magnetoresistance and Fermi surface in the Weyl semimetal WTe$_{2}$. We discovered a surprising feature: while our DFT+U+SOC calculations suggest a trivial Berry phase for electron hole pockets, the Lifshitz-Kosevich (LK) formula analysis indicates a non-trivial Berry phase for electron pockets. We attribute this inconsistency to the potential Zeeman effect arising from a large $g$-factor, a particularly relevant consideration for topological semimetals with strong spin-orbit coupling. Notably, prior works often neglect the $g$-factor. In this work, we systematically analyze why the determination of Berry phase exclusively from SdH oscillations may yield unreliable results.

\section{Context and problem formulation}

The Lifshitz-Kosevich theory provides the fundamental description of quantum oscillations, where the oscillatory component of magnetoconductivity is given by expression \cite{Zhao31122022,LIFSHITS19581,lifshitz1958theory,mallick2020piphasedifferencehall,shoenberg1984magnetic}
\begin{equation}
	\Delta\sigma_{xx}\propto\sqrt{\frac{B}{2F}}R_{S}R_{T}R_{D}\cos\left[2\pi\left(\frac{F}{B}- \frac{1}{2}+\beta+\delta\right)\right].
	\label{eq:oscillation}
\end{equation}
Here, $R_{T}$ describes the temperature dependence of the oscillation amplitude damping
\begin{equation}
	R_{T}=\frac{\lambda Tm^{*}/B}{\sinh(\lambda Tm^{*}/B)}.
	\label{eq:RT}
\end{equation}
The factor $R_{D}$ is the Dingle factor, describing the damping of oscillations with changing magnetic field
\begin{equation}
	R_{D}=\exp(-\lambda T_{D}m^{*}/B).
	\label{eq:RD}
\end{equation}
The parameter $\lambda$ in these expressions is defined as $\lambda=2\pi^{2}k_{B}/\hbar e$. In equation (\ref{eq:oscillation}) $\beta$ corresponds to the Berry phase $\phi_{B}=2\pi\beta$, while $\delta$ represents a geometry-dependent phase factor that vanishes for 2D systems but takes values of $\pm 1/8$ in 3D cases. Crucially, when $\phi_{B}=\pi$ ($\beta=0.5$) the oscillation phase should theoretically become zero for 2D systems and $\pm 1/8$ for 3D systems -- but only when all prefactors in equation (\ref{eq:oscillation}) remain positive. While $R_{T}$ and $R_{D}$ are strictly positive and thus cannot affect the phase, the spin factor $R_{S}$ introduces critical complications. This factor originates from spin degeneracy lifting in magnetic fields, which would normally produce two closely-spaced oscillation frequencies and beating patterns. As a consequence, this leads to an additional multiplier
\begin{equation}
	R_{S}=\cos\left(\pi\frac{E_{S}}{\Delta E}\right),
	\label{eq:RS}
\end{equation}
where $E_{S}$ is the spin splitting energy, and $\Delta E=\hbar\omega_{c}=\hbar eB/m^{*}$ is the Landau level spacing. If $E_{S}$ is not linear in magnetic field, $R_{S}$ may depend on the magnetic field, which leads to beating. However, if the spin splitting occurs due to the Zeeman effect, then $E_{S}=g\mu_{B}B=gBe\hbar/2m_{0}$, where $g$ is the Land\'e factor. Consequently,
\begin{equation}
	R_{S}=\cos\left(\pi\frac{gm^{*}}{2m_{0}}\right)
	\label{eq:RS_zeeman}
\end{equation}
becomes independent of the magnetic field. In other words, the beating disappears and only one oscillation frequency remains.

Thus, it is clear that $R_{S}$ is also an oscillating term whose sign depends on both $m^{*}$ and $g$. Using (\ref{eq:RT}), $m^{*}$ can be determined if the temperature dependence of the oscillation amplitude is known, but the $g$-factor cannot be directly and unambiguously determined from standard transport measurements. This term can also be easily understood as follows. Considering only the oscillating factor and using the Onsager relation $F = \hbar k_{F}^{2}/2e$ and the expression for Fermi energy $E_{F} = \hbar^{2}k_{F}^{2}/2m^{*}$, we obtain $F = m^{*}E_{F}/e\hbar$. In the presence of the Zeeman effect, the oscillation frequency is written as $F_{\uparrow} = m^{*}(E_{F} + g\mu_{B}B/2)/e\hbar$ and $F_{\downarrow} = m^{*}(E_{F} - g\mu_{B}B/2)/e\hbar$, from which we can write
\begin{align}
\Delta\sigma_{xx} &\propto \cos\left(2\pi\frac{F_{\uparrow}}{B}\right) + \cos\left(2\pi\frac{F_{\downarrow}}{B}\right) \nonumber \\
&\propto \cos\left(2\pi\frac{m^{*}(E_F + g\mu_B B/2)}{e \hbar B}\right) \nonumber \\
&\quad + \cos\left(2\pi\frac{m^{*}(E_F - g\mu_B B/2)}{e \hbar B}\right) \nonumber \\
&\propto 2\cos\left(\pi\frac{m^{*}g\mu_B}{e \hbar}\right)\cos\left(2\pi\frac{m^{*}E_F}{e \hbar B}\right) \nonumber \\
&\propto 2\cos\left(\pi\frac{gm^{*}}{2m_0}\right)\cos\left(2\pi\frac{F}{B}\right). 
\label{eq:zeeman_effect}
\end{align}
The first factor in the final expression corresponds to the spin factor introduced earlier (6). This shows that although the Zeeman effect should produce two different frequencies corresponding to carriers with different spins, only the original frequency remains, and an additional factor appears that depends on the effective mass and $g$-factor of the carriers.

\section{Contemporary approach to extracting quantum oscillation phase}

In practical studies, expression (2) is rarely used for determining oscillation phases, and in most cases, it is determined using what is known as a Landau level fan diagram constructed using the Lifshitz-Onsager quantization rule
\begin{equation}
	N = \frac{F}{B} + \gamma.
	\label{eq:landau_quantization}
\end{equation}
Here $\gamma = -\frac{1}{2} + \beta + \delta$ and let us set $\delta = 0$ for simplicity. Since Shubnikov-de Haas oscillations are caused by the oscillatory nature of the density of states $D(B)$, whenever a maximum of $D(B)$ is observed, a maximum of $\Delta\sigma_{xx}$ is observed too. Therefore, to determine the phase of oscillations $\gamma$, one should assign integer Landau level indices to maxima of $\Delta\sigma_{xx}$. As a result, we obtain the dependence $N = f(1/B)$, and by approximating it with a straight line, we can obtain the desired phase by finding the intersection point of this line with the $N$-axis.

This can be demonstrated using a simple qualitative model. Let us assume that Shubnikov-de Haas oscillations are observed from 0.15 to 1 T and are described by the equation
\begin{equation}
	\Delta\sigma_{xx} = A\exp\left(\frac{-D}{B}\right)\cos\left[2\pi\left(\frac{F}{B} - \frac{1}{2} + \beta\right)\right].
	\label{eq:shd_model}
\end{equation}
Let us set $A = 1$, $D = 0.5$, $F = 1$ T, and $\beta = 0.5$. As shown in Fig. 2, the straight line intersects the $N$-axis at 0, indeed, since $\gamma = -\frac{1}{2} + \frac{1}{2} = 0$. It is worth noting that most studies assign integer Landau level indices not to the maxima but to the minima of the oscillating $\Delta\sigma_{xx}$ component. In this case, the intercept reveals not $\gamma = -\frac{1}{2} + \beta$, but directly the Berry phase factor $\beta$ itself, since maxima and minima are shifted by $\pi$ relative to each other.

Here we would also like to note that in many modern studies of quantum oscillations and Berry phases, expressions similar to (2) are frequently used, but not for the oscillating part of magnetoconductivity $\Delta\sigma_{xx}$, but rather for magnetoresistivity $\Delta\rho_{xx}$ \cite{PhysRevLett.113.246402,PhysRevX.5.031023,doi:10.1073/pnas.1618004114,busch2018high,10.1063/5.0168564,Salawu_2021,wang2013large,Chen_2020,PhysRevB.93.121112,PhysRevB.95.085202,hu2016pi,PhysRevB.92.205134,salawu2022weak,Barua_2015,doi:10.1021/nn2024607,wu2019anomalous,Lonchakov_2019,Bobin_2019}. However, one must be careful here, since $\Delta\sigma_{xx}$ and $\Delta\rho_{xx}$ may either coincide in phase or not. This is due to the fact that in the general case $\sigma_{xx}$ is related to the resistivity tensor components as
\begin{equation}
	\sigma_{xx} = \frac{\rho_{xx}}{\rho_{xx}^{2} + \rho_{xy}^{2}}.
	\label{eq:conductivity_tensor}
\end{equation}
From this it is clear that if $\rho_{xx} \ll \rho_{xy}$, then $\sigma_{xx} \approx \rho_{xx}/\rho_{xy}^{2}$ and the corresponding oscillating parts $\Delta\sigma_{xx}$ and $\Delta\rho_{xx}$ coincide in phase. However, if $\rho_{xx} \gg \rho_{xy}$, then $\sigma_{xx} \approx 1/\rho_{xx}$ and $\Delta\sigma_{xx}$ is shifted in phase by $\pi$ compared to $\Delta\rho_{xx}$ \cite{PhysRevB.92.035123}. Therefore, when determining the phase of oscillations according to $\Delta\rho_{xx}$, it is important to consider the relationship between $\rho_{xx}$ and $\rho_{xy}$. However, in many topological insulators and semimetals where $\rho_{xx} \approx \rho_{xy}$, for correct analysis of oscillation phases, one must use the general expression \eqref{eq:conductivity_tensor}.

From this analysis, we can formulate a concise practical rule for the correct determination of the oscillation phase:
\begin{enumerate}
	\item If $\rho_{xx} \ll \rho_{xy}$, then $\Delta\rho_{xx}$ and $\Delta\sigma_{xx}$ are in phase. Consequently, to correctly determine the value of the phase factor $\gamma$, integer Landau level indices should be assigned to the \textbf{maxima} of either $\Delta\rho_{xx}$ or $\Delta\sigma_{xx}$.
	
	\item If $\rho_{xx} \gg \rho_{xy}$, then $\Delta\rho_{xx}$ and $\Delta\sigma_{xx}$ are phase-shifted by $\pi$. Consequently, to correctly determine the value of the phase factor $\gamma$, integer Landau level indices should be assigned to the \textbf{maxima} of $\Delta\sigma_{xx}$ or to the \textbf{minima} of $\Delta\rho_{xx}$.
	
	\item In the case where $\rho_{xx} \approx \rho_{xy}$, it is recommended to determine $\gamma$ \textbf{exclusively} from $\Delta\sigma_{xx}$.
\end{enumerate}

\begin{figure}
	\centering
	\includegraphics[width=0.8\columnwidth]{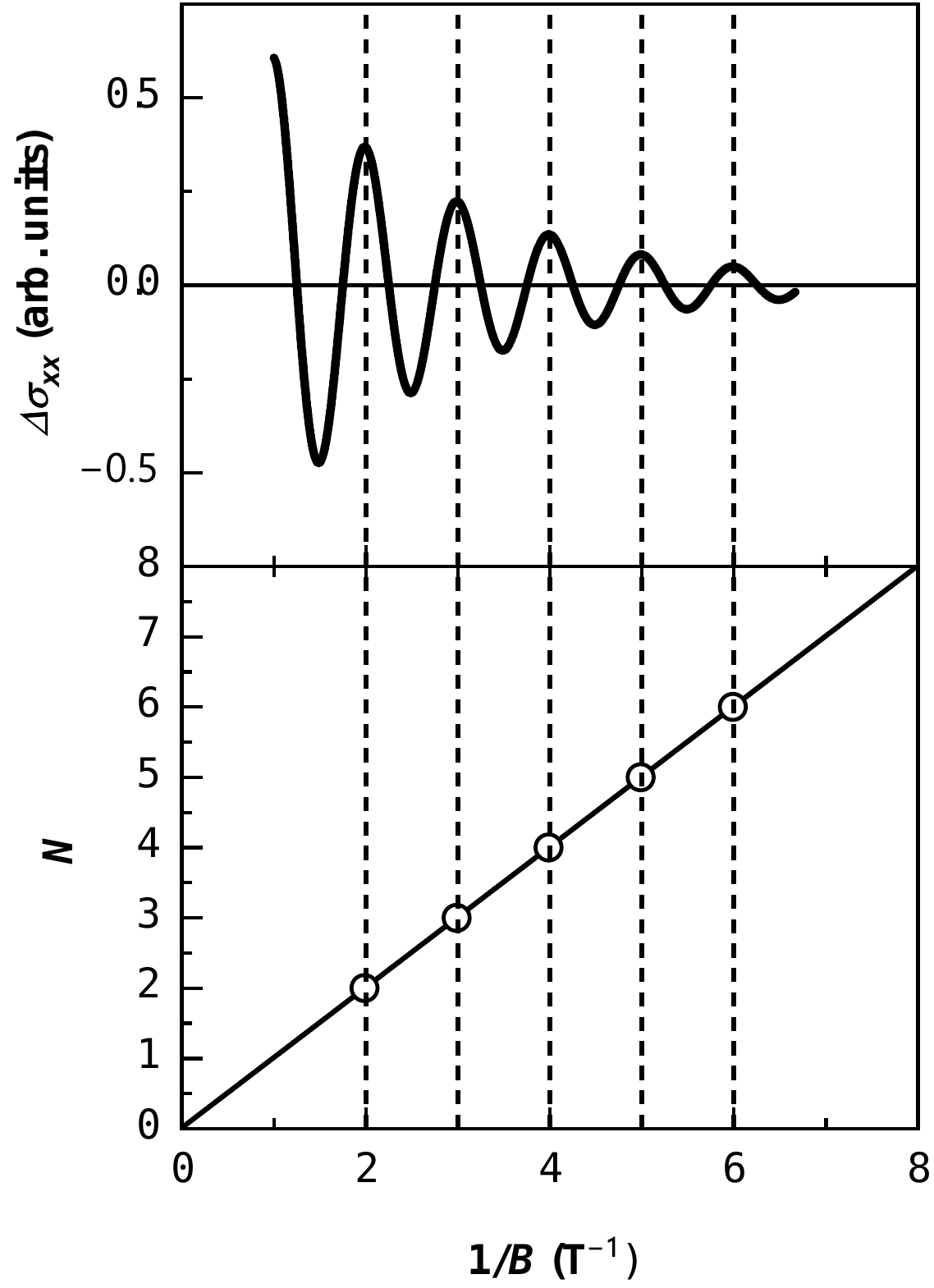}
	\caption{Oscillating part $\Delta\sigma_{xx}$ and corresponding Landau Level fan diagram for model (9).} 
	\label{fig2}
\end{figure}

\section{Interplay between the Zeeman effect and Berry phase}

In Shoenberg's 1984 book \cite{shoenberg1984magnetic}, it was noted that the determination of oscillation phases in earlier works \cite{doi:10.1098/rsta.1955.0007,PhysRev.95.1421,doi:10.1098/rspa.1939.0036,doi:10.1098/rsta.1952.0016} was used to reduce the possible values of the $g$-factor by half. Indeed, if we consider the dependence $R_{S}=f(g)$ for $m^{*}/m_{0}=0.25$ (Fig. 3), we can clearly see that for certain values of the $g$-factor $R_{S}>0$, while for others $R_{S}<0$. Interestingly, in that same year 1984, the seminal paper \cite{doi:10.1098/rspa.1984.0023} introducing the concept of geometric phase (later named Berry phase) was published. This means that during the analysis of oscillations in those earlier studies, the concept of Berry phase did not yet exist. At that time, if the intercept value gave $1/2$ (i.e., exactly matched the theoretical Onsager factor value of $1/2$), it indicated that $R_{S}>0$. On the contrary, an intercept of $0$ implied $R_{S}<0$, which immediately allowed eliminating half of possible $g$-factor values, as evident from Fig. 2. It's worth noting that in most modern studies determining Berry phase, the $R_{S}$ factor is not given proper attention and is often completely neglected. This is particularly relevant for topological materials with strong spin-orbit and electron-electron interactions, where $g$-factors can take widely varying values, up to giant ones $g = 2$--$100$ or ever more (see for example \cite{jiang2022giant,BEHERA202271,PhysRevB.107.155307}). Moreover, the possible presence of a non-trivial Berry phase makes it impossible to conclusively determine what causes the observed oscillation phase. This can be easily understood from the example discussed above. In that example, we assumed $R_{S}>0$ (i.e., $A > 0$) and set $\beta = 0.5$, resulting in zero oscillation phase. However, zero oscillation phase could also be obtained with zero Berry phase but with a $g$-factor yielding $R_{S}<0$ (i.e., $A < 0$), since 
$	-\cos\left[2\pi\left(\frac{F}{B}-\frac{1}{2}\right)\right]=\cos\left[\pi+2\pi\left(\frac{F}{B}-\frac{1}{2}\right)\right]=\cos\left[2\pi\left(\frac{F}{B}\right)\right].$
This clearly shows that the case of $R_{S}>0$ with $\beta = 0.5$ is equivalent to the case of $R_{S}<0$ with $\beta = 0$.

\begin{figure}
	\centering
	\includegraphics[width=0.8\columnwidth]{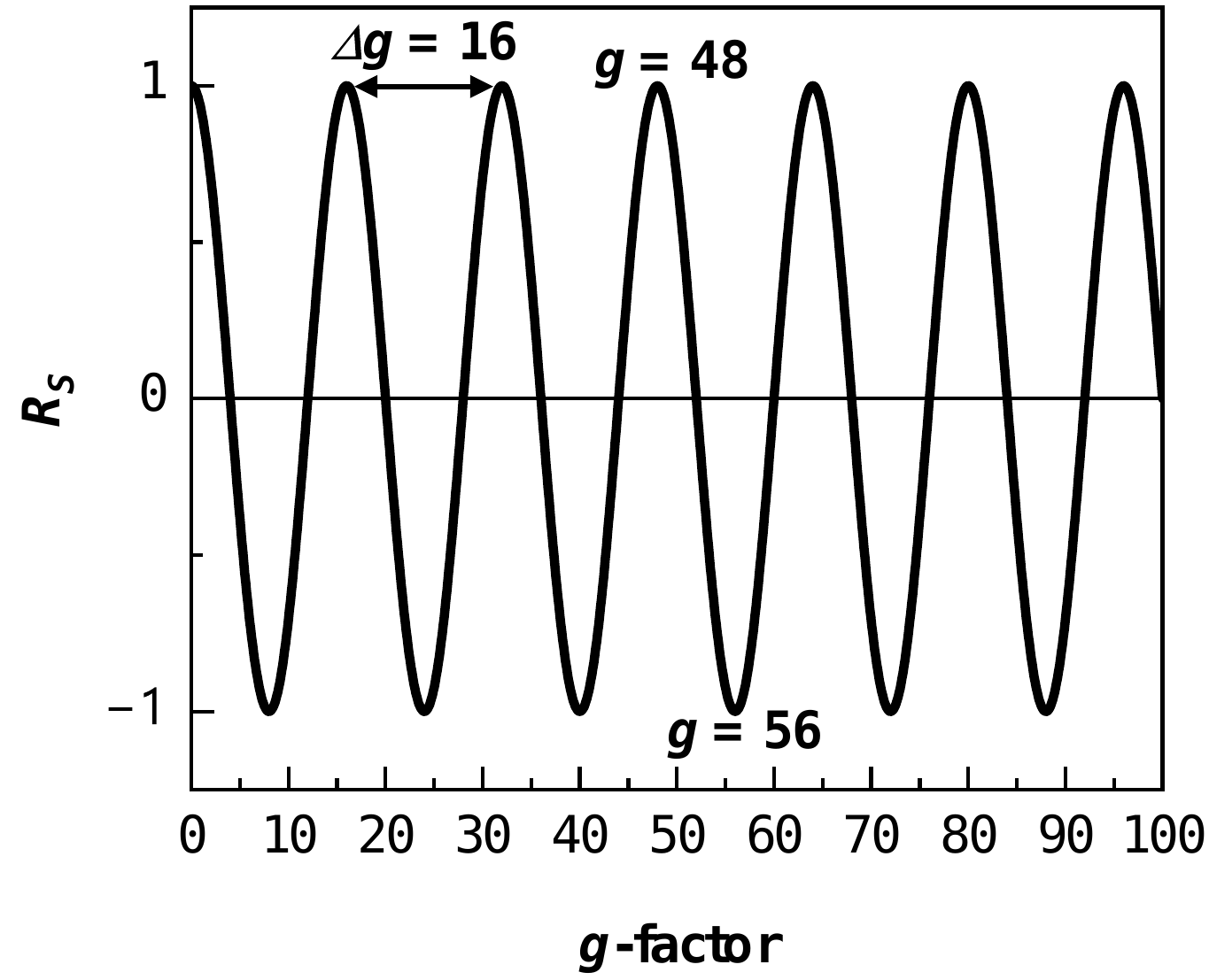}
	\caption{Dependence of $R_{S}=\cos\left(\pi\frac{gm^{*}}{2m_{0}}\right)$ on the $g$-factor at $m^{*}⁄m_{0} = 0.25$.} 
	\label{fig3}
\end{figure}

To illustrate this more clearly, let us consider the Hamiltonian for a 2D system that includes both linear and quadratic terms in the wave vector, as well as the Zeeman effect \cite{PhysRevB.84.035301,PhysRevB.87.085411,PismaZhETF.39.66,PhysRevB.97.195431,PhysRevB.82.085429,PhysRevLett.103.136803}:
\begin{equation}
	\hat{H}=v_{F}(\Pi_{x}\sigma_{y}-\Pi_{y}\sigma_{x})+\frac{\Pi^{2}}{2m^{*}}-\frac{1}{2}g\mu_{B}B\sigma_{z},
	\label{eq:hamiltonian}
\end{equation}
which yields the Landau levels
\begin{equation}
	E_{N}=\hbar\omega_{c}N \pm \sqrt{2\hbar v_{F}^{2}eBN+\left(\frac{1}{2}\hbar\omega_{c}-\frac{1}{2}g\mu_{B}B\right)^{2}}.
	\label{eq:landau_levels}
\end{equation}
This Hamiltonian is typical for topological insulators where $\Pi=\hbar\mathbf{k}+e\mathbf{A}$ with $\mathbf{A} = (0,Bx,0)$, $\sigma_{i}$ are the Pauli matrices, $m^{*}$ is the effective mass, $v_{F}$ is the Fermi velocity, $\mu_{B}$ is the Bohr magneton, and $g$ is the Land\'{e} $g$-factor, “$+$” and “$-$” branches are for electrons and holes, respectively. We note that in most topological insulators, for instance, only a single oscillation frequency is observed, meaning that only one electron or hole branch dominates. Therefore, we consider the scenario where the Fermi level lies within the conduction band, and the valence band does not contribute to the oscillations (i.e., we select the “$+$” sign in Eq. (12)).

First, let us examine how the Landau levels are modified by the Zeeman effect for conventional electrons with Landau level energies $E_{N}=\hbar\omega_{c}\left(N+\frac{1}{2}\right) \pm \frac{1}{2}g\mu_{B}B$. For a qualitative understanding, we take specific physical parameters for WTe$_{2}$ \cite{FOMINYKH2025182966}: oscillation frequency $F = 129.5$ T, Fermi energy $E_{F} = 60$ meV, and effective mass $m^{*} = 0.25m_{0}$. We analyze the dependence of Landau level energies $E_{N}$ for $N = 15$--$21$ on the inverse magnetic field $1/B$, corresponding to different $g$-factors: $g = 48$ for $R_{S} = 1$, $g = 56$ for $R_{S} = -1$ and $g = 52$ for $R_{S} = 0$ (Fig. 3). The resulting $E_{N}=f(1/B)$ curves for these cases are shown in Fig. 4. For $R_{S} = 1$ and $R_{S} = -1$ the Zeeman effect causes the spin-split Landau levels to overlap due to the relations $E_{N}^{\uparrow}=E_{N+6}^{\downarrow}$ for $g = 48$ and $E_{N}^{\uparrow}=E_{N+7}^{\downarrow}$ for $g = 56$. In the case of $R_{S} = 1$, the split Landau levels retain their original positions, whereas for $R_{S} = -1$, they shift to midway between the unsplit levels (Fig. 4a, b). To construct the corresponding Landau level (LL) fan diagrams, we determine the inverse magnetic field $1/B$ for each $N$ as the intersection point of $E_N = f(1/B)$ with the Fermi level. Fig. 4d shows that for $R_S = 1$, the phase factor $\gamma = -1/2$ the same as for unsplit levels. In contrast, for $R_S = -1$, $\gamma = 0$, corresponding to a phase shift of $\pi$ in the oscillations (Fig. 4e, f). For $R_S = 0$ ($g = 52$) the spin-split Landau levels no longer overlap (Fig. 4c). While one might expect oscillations from carriers with opposite spins, the LL fan diagram reveals that their phase factors $\gamma$ differ by 0.5 (Fig. 4e), equivalent to a $\pi$ phase difference. From the perspective of oscillations, this results in the sum of two identical oscillatory components with opposite signs, leading to complete cancellation, i.e.\ resulting $\Delta\sigma_{xx} = 0$ (Fig. 4f).

\begin{figure*}[!t]
	\centering
	\includegraphics[width=2\columnwidth]{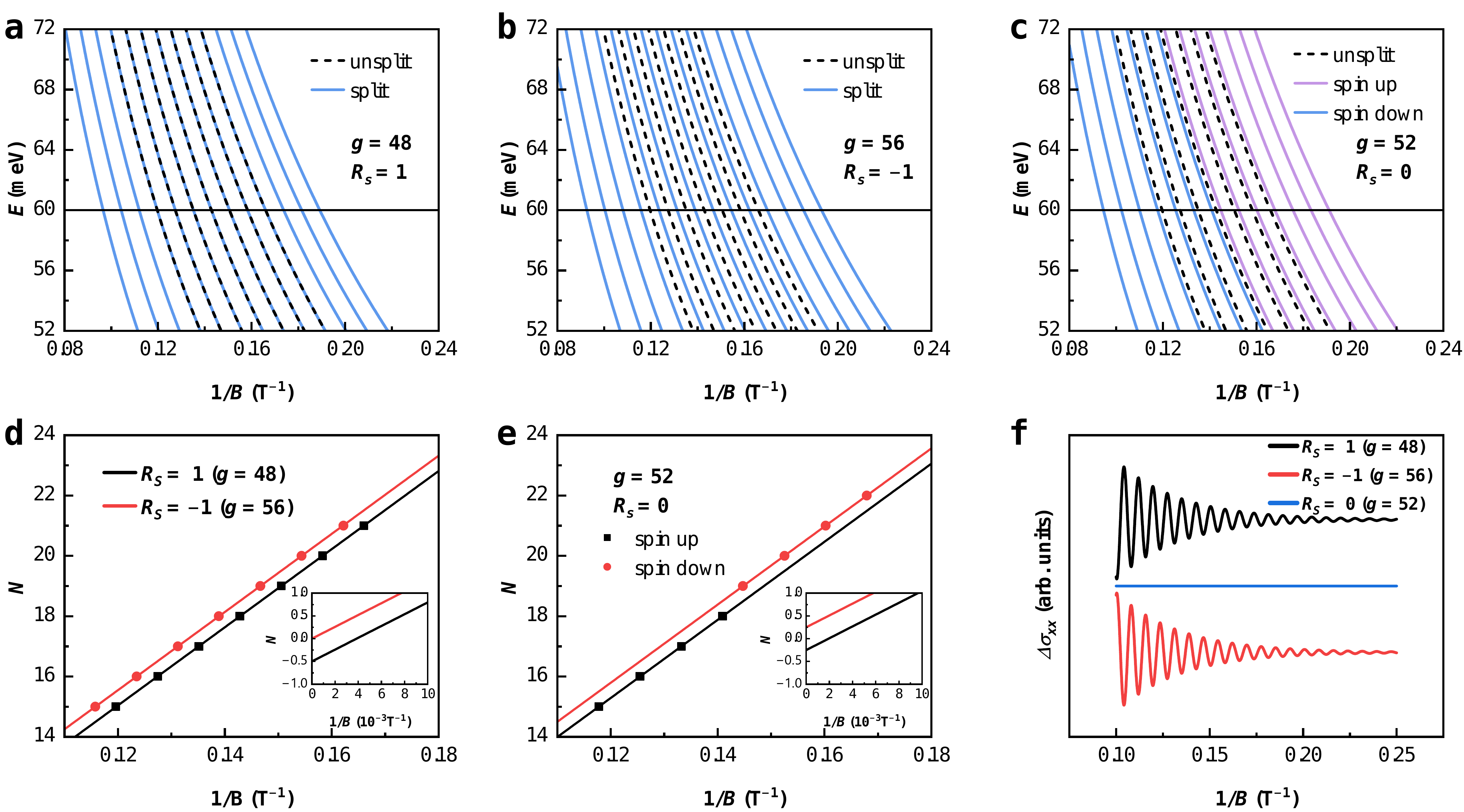}
	\caption{The influence of the Zeeman effect on Landau levels, LL fan diagrams, and oscillations for carriers with a quadratic dispersion relation for Landau levels $N = 15$--$21$ and specific physical parameters: oscillation frequency $F = 129.5$ T, Fermi energy $E_{F} = 60$ meV, and effective mass $m^{*} = 0.25m_{0}$. (a-c) Energies of the unsplit and spin-split Landau levels as a function of $1/B$ for three cases: (a) $R_s = 1$ ($g = 48$): the split levels coincide in position with the unsplit ones; (b) $R_s = -1$ ($g = 56$): the split levels are shifted exactly midway between the unsplit ones; (c) $R_s = 0$ ($g = 52$): the levels do not overlap, leading to the suppression of oscillations. (d-f) The corresponding LL fan diagrams and the oscillating component $\Delta\sigma_{xx}$. Case (b) demonstrates how a strong Zeeman effect ($R_s = -1$) can create a $\pi$ phase shift, precisely mimicking the signal of a non-trivial Berry phase ($\beta = 0.5$) in case (a).} 
	\label{fig4}
\end{figure*}

Taking into account the above, we can propose an extended Landau level fan diagram for cases where the Zeeman effect causes overlapping of split Landau levels, resulting only in a shift of the original levels
\begin{equation}
	N = \frac{F}{B} + \gamma + \frac{1}{2}\left(\frac{gm^{*}}{2m_{0}}\bmod 2\right).
	\label{eq:extended_fan}
\end{equation}
This expression adds an additional term $\frac{1}{2}\left(\frac{gm^{*}}{2m_{0}}\bmod 2\right)$ compared to the original \eqref{eq:landau_quantization} and depends on both effective mass and $g$-factor. This can be intuitively understood as follows. To determine which spin-split Landau levels overlap, we need to solve equation
\begin{equation}
	\begin{split}
		E_{N}^{\uparrow} &= E_{N+\Delta N}^{\downarrow} \\
		\hbar\omega_{c}\left(N + \frac{1}{2}\right) + \frac{1}{2}g\mu_{B}B &= \hbar\omega_{c}\left(N + \Delta N + \frac{1}{2}\right) - \frac{1}{2}g\mu_{B}B \\
		\Delta N &= \frac{g\mu_{B}B}{\hbar\omega_{c}} = \frac{E_{S}}{\Delta E} = \frac{gm^{*}}{2m_{0}}.
	\end{split}
	\label{eq:level_overlap}
\end{equation}
When $\Delta N = \frac{gm^{*}}{2m_{0}}$ takes even values, the split Landau levels occupy the same positions as the original ones, meaning the LL fan diagram remains unchanged and $\frac{1}{2}\left(\frac{gm^{*}}{2m_{0}}\bmod 2\right) = 0$. When $\Delta N = \frac{gm^{*}}{2m_{0}}$ is odd, this corresponds to split Landau levels lying midway between the unsplit ones, causing a $\pi$ phase shift in oscillations and consequently $\frac{1}{2}\left(\frac{gm^{*}}{2m_{0}}\bmod 2\right) = \frac{1}{2}$.

For fermions with linear dispersion, the situation becomes more complex because $\gamma$ becomes magnetic-field dependent, and the corresponding LL fan diagram deviates from a straight line. Let us first consider unsplit Landau levels with energies $E_N = \sqrt{2\hbar v_F^2 eBN}$, using the same parameters $F = 129.5$\,T, $E_F = 60$\,meV, and $v_F = 1.453 \times 10^5$\,m/s. As shown in Fig. 5a, although the $E_N = f(1/B)$ dependences differ qualitatively in this case, they intersect the Fermi level at the same points as the split levels for conventional electrons with quadratic dispersion and $R_S = -1$, meaning their oscillation phases coincide. When including the Zeeman effect, we obtain
\begin{equation}
	E_N = \sqrt{2\hbar v_F^2 eBN + \left(\frac{1}{2}g\mu_B B\right)^2}.
	\label{eq:linear_zeeman}
\end{equation}
Using \eqref{eq:linear_zeeman} and the conditions $E_F = \hbar k_F v_F$ and $E_N(B_N) = E_F$, we find
\begin{equation}
	N = \frac{F}{B} - \frac{1}{8}\frac{g^2\mu_B^2 B}{e\hbar v_F^2}.
\end{equation}
Comparison with \eqref{eq:landau_quantization} shows that $\gamma = -\frac{1}{8}\frac{g^2\mu_B^2 B}{e\hbar v_F^2}$. Fig.~\ref{fig4}b presents LL fan diagrams for different $v_F$ values at $g = 56$, while Fig.~\ref{fig4}c shows the corresponding dependences of the phase factor $\gamma$ on Landau level index $N$. Notably, $\gamma$ shows significant deviation from the ideal $\gamma = 0$ value for linear dispersion at small $N$, but approaches zero for large $v_F$ and large $N$. This demonstrates that in the large-$N$ limit, the phase factors for fermions with linear dispersion coincide with those for quadratic dispersion at $R_S = -1$. In Table 1, we have summarized our discussion on the relationship between the sign of $R_S$ and the Berry phase.

\begin{figure}
	\centering
	\includegraphics[width=1\columnwidth]{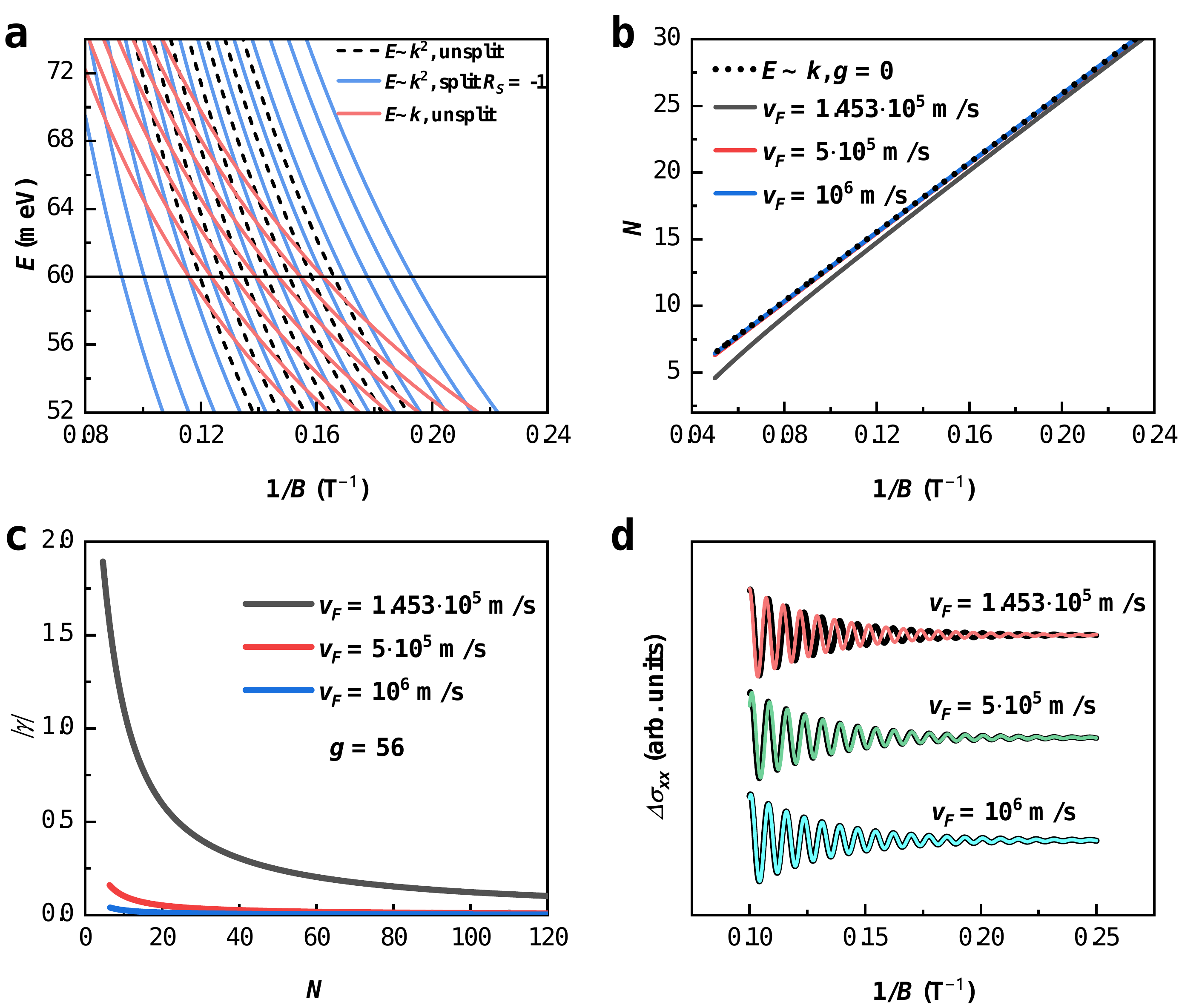}
	\caption{ (a) Dependencies of energies for unsplit and split Landau levels with indices $N = 15$--$21$ on inverse magnetic field for fermions with quadratic dispersion at $R_S = -1$, as well as for unsplit Landau levels of fermions with linear dispersion. 
		(b, c) LL fan diagram and dependence of phase factor $\gamma$ on LL index for fermions with $E \sim k$, $g = 56$ and different $v_F$ values. 
		(d) Corresponding oscillating components $\Delta\sigma_{xx}$, where black curves represent $\Delta\sigma_{xx}$ for unsplit Landau levels, while colored curves show split cases with different $v_F$.} 
	\label{fig5}
\end{figure}

\begin{table}[t]
	\centering
	\caption{Relationship between $R_S$ sign, Berry phase $\beta$, and observed oscillation phase.
	\textit{Note:} $\beta$ is Berry phase factor; $\delta$ is geometry phase factor ($0$ for 2D, $\pm 1/8$ for 3D).}
	\label{table:RS_berry_ambiguity}
	\footnotesize
	\begin{tabular}{ccccc}
		\hline
		\textbf{Case} & \boldmath$R_S$ & \boldmath$\beta$ & \boldmath$\gamma$ & \textbf{Interpretation} \\
		& \textbf{sign} & & & \textbf{(if $R_S$ ignored)} \\
		\hline
		1 & $+$ & 0 & $-1/2 + \delta$ & Correct: trivial \\
		2 & $+$ & 1/2 & $0 + \delta$ & Correct: nontrivial \\
		3 & $-$ & 0 & $0 + \delta$ & \textbf{False: nontrivial} \\
		4 & $-$ & 1/2 & $-1/2 + \delta$ & \textbf{False: trivial} \\
		\hline
		
	\end{tabular}
\end{table}

It should be noted that for topological insulators, knowing the $g$-factor may not be necessary for correct determination of oscillation phase. This is because in TIs, quantum oscillations can originate either from bulk states with zero Berry phase or surface states with $\pi$ Berry phase. Bulk and surface states can be effectively separated by studying magnetoresistance in tilted magnetic fields. For surface states, the oscillation frequency dependence should follow the expression for 2D Fermi surface
$F(\theta) = {F_0}/{\cos\theta}$ \cite{PhysRevB.86.045131,PhysRevB.82.241306}, while for 3D states the $F(\theta)$ dependence won't follow this expression. For example, in TI Bi$_2$Se$_3$, when oscillations come from bulk states, $F(\theta)$ can be described by the expression for 3D spheroidal Fermi surface
$	F(\theta) = F_0\left(\cos^2\theta + \left({k_{\mathrm{F}}^{x}} / {k_{\mathrm{F}}^{z}}\right)^2 \sin^2\theta\right)^{-1/2}$ \cite{busch2018high,PhysRevB.88.195107}.
However, the situation becomes much more complicated in topological semimetals, since Dirac or Weyl fermions with nontrivial Berry phase exist in the bulk of the sample. In this case, studying magnetoresistance in tilted magnetic fields won't allow unambiguous determination of whether oscillations come from topologically trivial or non-trivial carriers. The situation becomes especially complicated when oscillations contain multiple frequencies corresponding to different Fermi surface sheets, which is particularly relevant for topological semimetals. Therefore, for correct determination of oscillation phase, it's necessary to know the $g$-factor for each carrier type corresponding to its Fermi surface sheet.

Recently, the so-called spin-zero effect has been actively studied, where the oscillation amplitude vanishes at certain $g$-factor values due to $R_S = 0$. This effect is typically detected by examining field-dependent magnetoresistance or spin torque at various angles between the magnetic field direction and the axis perpendicular to the sample plane. The spin-zero effect has been observed in Te nanoribbons \cite{PhysRevB.108.245409}, the kagome metal CsV$_3$Sb$_5$ \cite{chen2023magnetic}, and topological semimetals ZrSiS \cite{PhysRevB.96.045127}, ZrTe$_5$ \cite{doi:10.1073/pnas.1804958115,liu2016zeeman}, CaAgAs \cite{PhysRevResearch.2.012055} and WTe$_2$ \cite{Bi_2018}. For instance, in \cite{doi:10.1073/pnas.1804958115} it was found that at a specific angle ($\theta \approx 84^\circ$), the magnetoresistance oscillation amplitude in ZrTe$_5$ drops to zero before reappearing. Landau level fan diagram analysis revealed an abrupt $\pi$ phase shift at $\theta \approx 84^\circ$, clearly indicating the $g$-factor transition between positive and negative (or vice versa) $R_S$ regions through zero. Interestingly, while the authors of \cite{doi:10.1073/pnas.1804958115} claim $\phi_B = \pi$ for $\theta < 84^\circ$ (implying $R_S > 0$), there's no fundamental reason to exclude the possibility of $R_S < 0$ at $\theta < 84^\circ$. The very existence of the spin-zero effect, demonstrating strong $g$-factor anisotropy, makes it impossible to unambiguously determine the sign of $R_S$ in perpendicular fields.

Another notable result appears in \cite{SALAWU2022116079}, where quantum oscillations in Bi$_{0.97}$Sb$_{0.03}$ were studied under various magnetic field (up to 50\,T) and current configurations, extending to the quantum limit. The authors used an expression similar to (2) to fit the oscillating components $\Delta\rho_z$, $\Delta\rho_x$ and $\Delta\rho_y$ and determine the oscillation phase. Their fitting indicated a nontrivial Berry phase. After extracting other parameters (effective mass, Dingle temperature), they treated the $g$-factor in (2) as a fitting parameter, obtaining large values ($\sim$70 for $\Delta\rho_z$ and $\Delta\rho_y$ and $\sim$40 for $\Delta\rho_x$). By rewriting the spin factor $R_S$ as \eqref{eq:RS}, they found $E_S/\Delta E \approx 2$, meaning Zeeman-split Landau levels coincide with unsplit level positions. Landau fan diagram analysis also suggested a nontrivial Berry phase. However, it's important to note that the authors determined the Berry phase from resistance oscillations without specifying whether the $\rho_{xx}/\rho_{xy}$ ratio was considered. Conversely, by assuming a nontrivial Berry phase, they constrained $E_S/\Delta E$ to even integers. Yet, with zero Berry phase, $E_S/\Delta E$ would be odd, and the resulting oscillation phases would coincide. This creates a contradiction that prevents unambiguous determination of both $R_S$ sign and Berry phase.

We must note that there exists an extremely limited number of studies explicitly demonstrating the scenario depicted in Fig.~\ref{fig4}b, which can produce a spurious Berry phase signature for conventional electrons. This scarcity is unsurprising, as identification through quantum oscillations alone is notoriously difficult. A significant result was obtained in works \cite{GUDINA2021167655,bogolubskii2021quantum,nano12071238,neverov2020effective,gudina2018electron,10.1063/1.4983183,10.1063/1.1935753} investigating HgTe-based quantum wells and heterostructures. 
When analyzing magnetoresistance quantum oscillations in HgTe quantum wells with inverted band spectra, the authors of \cite{GUDINA2021167655} observed an anomalous zero oscillation phase. This finding could mistakenly suggest a topological electronic structure (i.e., Dirac or Weyl fermions). However, as noted in \cite{GUDINA2021167655}, the band structure of HgTe quantum wells is trivial and devoid of topological features. In their theoretical and experimental study, the authors discovered a giant Land\'{e} factor ($g \approx 63$), indicating Zeeman splitting of Landau levels sufficiently strong to position the split levels midway between unsplit levels -- precisely as illustrated in Fig.~\ref{fig4}b. 
This configuration introduces an additional $\pi$-phase shift in quantum oscillations, creating the illusion of a Berry phase. These results reemphasize the critical importance of accounting for Zeeman effects when determining quantum oscillation phase.

Based on the study of Te nanowires presented in the article \cite{PhysRevB.108.245409}, it becomes particularly clear why accounting for the sign of the spin factor $R_S$ is critically important for the correct determination of the Berry phase. In this work, the authors observed a trivial phase (an intercept near $-1/8$) in the Landau level fan diagram at large magnetic field tilt angles, which seemingly indicated the absence of nontrivial topology. However, through angle-dependent measurements, a spin-zero effect was discovered, leading to an additional $\pi$ phase shift that masked the true nontrivial Berry phase ($\sim -5/8$), which manifested at other field orientations. This case vividly demonstrates that even in a material with a predicted and otherwise confirmed topological nature, the direct extraction of the Berry phase from quantum oscillations can lead to an erroneous "trivial" conclusion if the dependence of $R_S = \cos(\pi g m^*/2 m_0)$ on the effective mass and the $g$-factor is not considered. Thus, the work on Te underscores that neglecting the sign of $R_S$ does not merely introduce uncertainty but can systematically conceal nontrivial topology, making a combined experimental approach—including angle-dependent measurements and independent determination of the $g$-factor—essential for an unambiguous interpretation.

While our analysis has primarily focused on the ambiguity introduced by the Zeeman effect and the spin factor $R_S$, it is crucial to acknowledge that this is not the sole source of possible misinterpretation of the Berry phase. Even in the absence of Zeeman splitting, the intricate structure of Landau levels in topological semimetals can lead to anomalous phase shifts that deviate from the simple $\pm 1/8$ or $\pm 5/8$ values. A seminal study by Wang, Lu, and Shen \cite{PhysRevLett.117.077201} demonstrated that for Weyl and Dirac semimetals, the phase shift $\phi$ extracted from a Landau fan diagram can change non-monotonically as a function of Fermi energy $E_F$, reaching values beyond the conventional set (e.g., $\phi = -9/8$) near the Lifshitz transition point due to inter-Landau-band scattering and the complex, non-parabolic dispersion. Furthermore, they showed that topological band inversion alone can generate beating patterns in quantum oscillations, which could be mistaken for effects arising from Zeeman splitting. This work underscores a fundamental challenge: the phase $\gamma$ in the Lifshitz-Onsager quantization rule becomes a non-universal, energy-dependent parameter $\gamma(E_F)$ in topological semimetals, rather than a constant simply related to the Berry phase $\phi_B = 2\pi\beta$. Therefore, our present discussion on the Zeeman effect and the earlier work on band structure effects \cite{PhysRevLett.117.077201} collectively paint a more complete and cautionary picture: unambiguous determination of the Berry phase from quantum oscillations is fraught with difficulties, requiring extreme caution and complementary measurements to disentangle these various confounding mechanisms.

Finally, it is crucial to recognize that the impact of a large $g$-factor is not limited to phase shifts. As shown in recent studies \cite{PhysRevB.102.041204}, in topological semimetals with Dirac-like dispersion, a sufficiently large $g$-factor can enable a crossing and inversion of Landau levels at magnetic fields beyond the quantum limit. This leads to a new type of quantum oscillation that is periodic in $B$ (not $1/B$) with a period set by the ratio $g/v_F$ (not the Fermi surface area). This phenomenon underscores the profound and diverse ways in which the $g$-factor can manifest in quantum oscillations, transitioning from a source of interpretative ambiguity in the Lifshitz-Kosevich regime to the primary driver of entirely new oscillatory behavior in the ultra-quantum limit.

\section{Magnetic field dependence of the Fermi level}

\begin{figure}
	\centering
	\includegraphics[width=1\columnwidth]{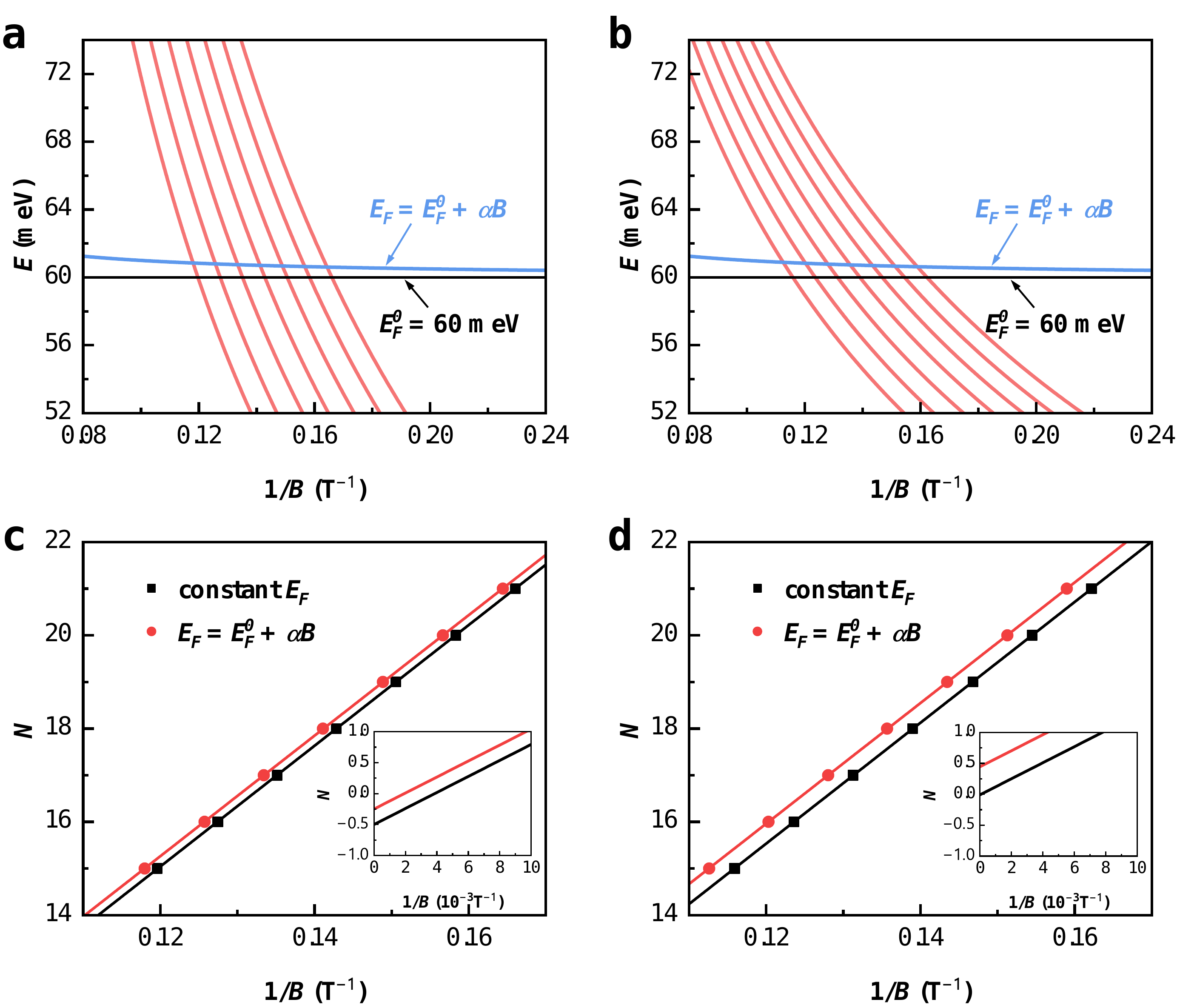}
	\caption{ The influence of the magnetic field dependence of the Fermi level, $E_F(B)$, on the LL fan diagram. (a, b) Energies of the unsplit Landau levels for carriers with (a) quadratic and (b) linear dispersion relations. The black horizontal line represents the constant Fermi level $E^0_F$. The blue inclined line represents the field-dependent Fermi level: $E_F = E^0_F + \alpha B$. (c, d) Corresponding LL fan diagrams. As can be seen, in both cases the $E_F(B)$ dependence leads to a phase shift of the oscillations, which can be mistaken for a non-trivial Berry phase in the trivial case and vice versa.} 
	\label{fig6}
\end{figure}

As noted in \cite{PhysRevB.97.195431}, another possible mechanism receiving significantly less attention than the Zeeman effect that can distort oscillation phases is the dependence of the Fermi level on the magnetic field. Here, we briefly discuss how this mechanism can substantially mask or mimic the Berry phase. A similar effect was demonstrated via tunneling spectroscopy in p-(Bi$_{1-x}$Sb$_{x}$)$_2$Te$_{3}$/n-InP device \cite{yoshimi2014dirac}, exhibiting a linear rate of $\alpha = 1$\,meV/T. 
Thus, we consider the simplest qualitative example where the Fermi level depends linearly on the magnetic field, $E_F = E_F^0 + \alpha B$. For our model, we take $E_F^0 = 60$\,meV and $\alpha = 0.1$\,meV/T and examine Landau levels (LLs) with indices $N = 15$--$21$. For clarity, we consider only non-split Landau levels. In the case of quadratic dispersion, the LL fan diagram is modified as:
\begin{equation}
	N = \frac{F}{B} - \frac{1}{2} + \frac{\alpha m^{*}}{\hbar e}.
	\label{eq:modified_fan}
\end{equation}
Figure~\ref{fig6}a shows the Landau levels for electrons with quadratic dispersion plotted against inverse magnetic field. The black line indicates a constant Fermi level, while the blue line represents the field-dependent Fermi level $E_F = E_F^0 + \alpha B$. Figure~\ref{fig6}c displays the corresponding LL fan diagram, revealing a shift in the intercept by approximately $\frac{\alpha m^{*}}{\hbar e}$. 
Notably, in many topological insulators, the intercept deviates from the ideal value of 0.5 \cite{doi:10.1126/science.1189792,PhysRevB.84.035301,PhysRevB.82.241306}. This is often attributed to Berry phase distortion caused by non-ideal Dirac dispersion or the Zeeman effect. However, as we demonstrate here, such non-ideal intercepts can also arise in ordinary fermion systems with a magnetic-field dependent Fermi level.

For the linear dispersion relation, the LL fan diagram takes the following form:
\begin{equation}
N=\frac{F}{B}+\frac{\alpha^{2}B}{2e\hbar v_{F}^{2}}+\frac{\alpha E_F^0}{e\hbar v_{F}^{2}}.
\label{eq:linear_fan}
\end{equation}
Interestingly, in this case, an additional linear term appears. However, it has an almost negligible effect on the LL fan diagram, as this term is more than two orders of magnitude smaller than the other contributions. This leads to an important conclusion: the dependence $E_{F}(B)$ may not alter the oscillation frequency which significantly complicates the identification of this effect from the experiment. It is worth noting that a similar form of the LL fan diagram was theoretically derived by the authors of \cite{PhysRevB.87.085411} for topological insulators with broken particle-hole symmetry: $N -\Lambda = B_{0}/B + A_{1} + A_{2}B$. 
Fig.~\ref{fig6}d shows the corresponding LL fan diagram, where it can be seen that the magnetic-field dependence of the Fermi level in this case leads to a significant shift in the intercept, bringing it close to $0.5$. In a standard analysis of quantum oscillations, such a shift could be mistakenly attributed to a trivial Berry phase.

When accounting for the Zeeman effect as well, the LL fan diagrams will take the following generalized forms:

\begin{equation}
	N = \frac{F}{B} - \frac{1}{2} + \frac{1}{2} \left( \frac{g m^*}{2 m_0} \bmod 2 \right)  + \frac{\alpha m^*}{\hbar e}
	\label{eq:quadratic_extended}
\end{equation}
for quadratic dispersion, and
\begin{equation}
	N = \frac{F}{B} - \frac{1}{8} \frac{g^2 \mu_B^2 B}{e \hbar v_F^2} + \frac{\alpha^2 B}{2 e \hbar v_F^2} + \frac{\alpha E_F^0}{e \hbar v_F^2}
	\label{eq:linear_extended}
\end{equation}
for linear dispersion relations.

Thus, the magnetic-field dependence of the Fermi level is another crucial mechanism that can obscure or mimic the Berry phase. This mechanism can significantly complicate the extraction of the oscillation phase, as it cannot be identified using quantum oscillations alone.

While the linear dependence $E_F = E_F^0 + \alpha B$ serves as a useful pedagogical example, a more universal and fundamental scenario arises in systems with fixed carrier density $n_e$. In this case, the Fermi level is not constant but must oscillate with magnetic field to conserve the number of states below it \cite{shoenberg1984magnetic}. As the magnetic field increases, Landau levels pass through the Fermi energy, causing $E_F$ to oscillate with a phase that is shifted by $\pi/2$ relative to the oscillations in the density of states $D(E)$ (and hence, relative to $\Delta\rho_{xx}$ or $\Delta\sigma_{xx}$) since $n_e = \int_0^{E_F} D(E)\,dE$.

These oscillations of $E_F$ directly impact the LL fan diagram. The quantization condition becomes
\begin{equation}
S(E_F, B) = \frac{2\pi eB}{\hbar}(N+\gamma),
\end{equation}
where $S(E_F, B)$ is the cross-sectional area of the Fermi surface at energy $E_F$. This leads to an additional oscillatory term in the LL fan diagram equation
\begin{equation}
	N = \frac{F}{B} + \gamma + \tilde{\gamma}(B),
\end{equation}
where $\tilde{\gamma}(B)$ is an oscillatory correction with the same period as the quantum oscillations themselves. Consequently, when constructing a LL fan diagram by assigning integer indices $N$ to maxima or minima of $\Delta\rho_{xx}$, one effectively plots $N$ vs $1/B$ not along a straight line, but along a curve that oscillates around the ideal line. This can lead to a significant dispersion of data points and systematic errors in determining the intercept $\gamma$ from a linear fit, especially if the fit is performed over a limited range of indices or magnetic fields. In extreme cases, the phase extracted from a naive linear fit could even appear to be shifted from its true value. This effect provides another mechanism that can mask or mimic a non-trivial Berry phase, and it is fundamentally unavoidable in any quantum oscillation experiment where the carrier density is fixed.

It is worth noting that we were unable to find studies that explicitly demonstrate the impact of Fermi energy oscillations on the phase of quantum oscillations. Although the influence of $E_F$ oscillations has been investigated in two-dimensional materials, their detailed effect on the oscillation phase has not been thoroughly examined \cite{BARNES1996608,minkov2024manifestationfermileveloscillations}.

 Furthermore, another significant source of ambiguity arises from the non-parabolicity of the electron dispersion. Within the parabolic approximation, the effective mass $m^* = \frac{\hbar^2}{2\pi} \frac{\partial S}{\partial E}$ extracted from the temperature damping of oscillations (Eq. 3) is assumed to be constant. However, in topological materials the dispersion may deviate from ideal parabolic or linear behavior \cite{PhysRevB.84.035301}. In this case, the effective mass becomes energy-dependent $m^*(E)$. Since the spin factor $R_S$ (Eq. 6) critically depends on $m^*$, its sign and magnitude can change with shifts in the Fermi level $E_F$. This creates an additional mechanism capable of mimicking a phase shift attributed to the Berry phase. Moreover, the phase factor $\gamma$ itself in the Lifshitz-Onsager quantization rule ceases to be a universal constant in the case of strong non-parabolicity and can become a function of the magnetic field and Landau level index $N$, taking on anomalous values \cite{PhysRevLett.117.077201}, which further complicates the unambiguous extraction of the geometric Berry phase.
 
 In light of the discussed difficulties, an important alternative to phase analysis has emerged, based on the temperature dependence of the oscillation frequency $F(T)$, which allows for the identification of topological carriers through their linear dispersion relation. For a quadratic dispersion relation $S = \frac{2\pi m^* E_F}{\hbar^2}$, and it is clear that $\frac{\partial m^*}{\partial E} = 0$. In contrast, for a linear dispersion relation $S = \frac{\pi E_F^2}{\hbar^2 v_F^2}$, we obtain $m^* = \frac{E_F}{v_F^2}$ and thus $\frac{1}{m^*}\frac{\partial m^*}{\partial E} = \frac{1}{E_F}$. As demonstrated by Guo et al.~\cite{guo2021temperature}, such a non-zero correction leads to an additional contribution to the temperature dependence of the oscillation frequency \cite{guo2021temperature}:
 	\begin{equation}
 		F(T) = F_0 - \Theta \frac{(\pi k_B T)^2}{\beta^2 F_0(E_F)} ,
 	\end{equation}
 	where $\beta = e\hbar/(2m^*)$, and the dimensionless coefficient $\Theta$ takes characteristic values depending on the carrier type: $\Theta = 0$ and $1/16$ for a quadratic and linear spectrum in a high-carrier-density metal, respectively, $\Theta = 5/48$ for a Dirac semimetal, and $\Theta = 1/48$ for a trivial parabolic spectrum. This approach, which does not require knowledge of the $g$-factor and has been verified on materials such as Cd$_3$As$_2$ (topological) and Bi$_2$O$_2$Se (trivial), avoids the ambiguities inherent in Berry phase analysis and represents a powerful independent tool for diagnosing topological Fermi surfaces \cite{guo2021temperature}.

\section{Accounting for the orbital moment}

The modern semiclassical theory of a Bloch electron in a magnetic field, developed in recent decades, establishes that the total phase $\lambda$ entering the Lifshitz-Onsager quantization rule is a sum of three distinct contributions: $\lambda = \phi_B + \phi_R + \phi_Z$ \cite{PhysRevX.8.011027}. Here, $\phi_B$ is the Berry phase, $\phi_Z$ is the contribution from the Zeeman energy splitting, and $\phi_R$ is the phase arising from the orbital magnetic moment of the Bloch wave packet. The resulting quantization rule can be expressed as \cite{PhysRevX.8.011027,annurev021331,PhysRevB.53.7010,RevModPhys.82.1959}
\begin{equation}
	l^2 S + \lambda = 2\pi N + \phi_M,
	\label{eq:alex_quantization}
\end{equation}
where $l = \sqrt{\hbar / eB}$ is magnetic length, $S$ is the Fermi surface cross-section in $\mathbf{k}$-space and $\phi_M$ is the Maslov correction which equals $\pi$ for orbits that are deformable to a circle \cite{KELLER1958180}, but vanishes for a figure-of-eight orbit \cite{PhysRevLett.119.256601}.
 The emergence of the $\phi_R$ phase has a profound physical origin: as the wave packet traverses a closed orbit in $\mathbf{k}$-space, its intrinsic orbital magnetic moment interacting with the external magnetic field leads to an additional dynamical phase accumulation. This contribution can be significant in materials with strong spin-orbit coupling and complex band structures, where the magnitude of orbital moment is large. Consequently, the experimentally observed phase offset in a Landau level fan diagram or in quantum oscillations corresponds to the total phase $\lambda$, not solely to the Berry phase $\phi_B$.

The present work has focused on analyzing the ambiguity arising from the interplay between the Berry phase $\phi_B$ and the Zeeman contribution $\phi_Z$ (which is closely related to the spin factor $R_S$ in the Lifshitz-Kosevich formalism), under the initial assumption that the orbital moment contribution $\phi_R$ is small. However, in the general case, especially for topological materials, the contribution of $\phi_R$ can be substantial and must be accounted for in a complete quantitative analysis. Incorporating the orbital moment does not invalidate but rather reinforces the central conclusion of our work: the unambiguous determination of the true Berry phase $\phi_B$ from quantum oscillation data alone remains a fundamentally challenging task. The problem of decomposition is now further exacerbated: even if one could somehow accurately determine the $\phi_Z$ contribution, extracting $\phi_B$ would additionally require knowledge of $\phi_R$. For instance, an observed total phase of $\lambda = \pi$ could stem from three fundamentally different scenarios: (i) nontrivial topology ($\phi_B = \pi$) with negligible $\phi_R$ and $\phi_Z$; (ii) trivial topology ($\phi_B = 0$) compensated by a large positive $\phi_R$ contribution; or (iii) trivial topology and a small $\phi_R$, but with a large $g$-factor leading to an effective phase shift via $R_S$, as discussed in detail in previous sections.

The symmetry analysis performed by Alexandradinata et al. \cite{PhysRevX.8.011027} demonstrates that for certain (magnetic) space groups and specific field orientations, the total phase $\lambda$ (or the sum of phases over degenerate subbands) can be quantized and protected by symmetry, becoming a topological invariant of magnetotransport. In these special cases, the problem of ambiguity is lifted. However, for arbitrary field orientations and specific space groups such protection may absent. Therefore, for the correct determination of the topological Berry phase $\phi_B$ in the general case, independent experimentation or first-principles calculations are required to estimate the contributions from both the orbital moment $\phi_R$ and the $g$-factor. This further underscores the necessity of the combined experimental approach and calls for further theoretical efforts aimed at developing practical methods for disentangling these three contributions in experiment.

\section{Conclusion}
In this work, we have systematically analyzed the fundamental challenges and ambiguities in extracting the Berry phase $\phi_B$ solely from quantum oscillation measurements, such as the Shubnikov-de Haas effect. While the Berry phase remains a crucial indicator of band topology, its experimental determination is often obscured by several factors that can mimic or mask its signature.

We have demonstrated that the spin factor $R_S = \cos(\pi g m^* / 2 m_0)$, governed by the often-unknown Landé $g$-factor, can lead to a $\pi$ phase shift in quantum oscillations. This effect can make a topologically trivial system ($\phi_B = 0$) appear non-trivial, and vice-versa, if $R_S$ is neglected. This ambiguity is particularly acute in topological materials with strong spin-orbit coupling, where $g$-factors can be large and widely varying.

Furthermore, we have highlighted the role of another often-overlooked mechanism: the magnetic field dependence of the Fermi level $E_F(B)$. This dependence can also lead to significant shifts in the Landau level fan diagram intercept, potentially mimicking a non-trivial Berry phase even in ordinary Fermion systems.

Most importantly, following the modern semiclassical theory \cite{PhysRevX.8.011027}, we have emphasized that the experimentally accessible phase in the Lifshitz-Onsager rule is the total phase $\lambda = \phi_B + \phi_R + \phi_Z$, which includes contributions from the Berry phase $\phi_B$, the orbital magnetic moment $\phi_R$, and the Zeeman effect $\phi_Z$. This tripartite nature of the phase makes the unambiguous extraction of $\phi_B$ a formidable decomposition problem without additional constraints.

Therefore, we conclude that determining both the $g$-factor and the Berry phase from quantum oscillations alone is fundamentally unreliable in the general case. A zero phase offset in a Landau fan diagram is not a smoking gun for Dirac or Weyl physics, as it can be equally explained by a combination of a trivial Berry phase and a negative $R_S$, or influenced by $E_F(B)$ and $\phi_R$.

To overcome these ambiguities, a combined experimental approach is essential. We advocate for:
\begin{enumerate}[label=$\bullet$]
	\item The independent determination of the $g$-factor via techniques like magneto-infrared spectroscopy or the spin-zero effect.

	\item The use of this fixed $g$-factor value to constrain the analysis of quantum oscillations within the Lifshitz-Kosevich formalism.

	\item Where possible, use symmetry considerations \cite{PhysRevX.8.011027} or first-principles calculations to estimate the contribution of the orbital moment $\phi_R$.
	
\end{enumerate}

Additionally, the study \cite{guo2021temperature} proposed a universal method based on measuring the temperature dependence of the oscillation amplitude. This is a highly versatile technique, as it only requires high-precision measurements of the quantum oscillations at different temperatures.

This direct constraint breaks the degeneracy between a negative $R_S$ and a non-trivial $\phi_B$, paving the way for a more robust identification of topological states. We hope this discussion will stimulate further research to refine the interpretation of quantum oscillations and develop robust, multi-faceted protocols for determining the Berry phase in real materials.

\section*{Acknowledgements}

The work was carried out within the framework of the state assignment of the Ministry of Science and Higher Education of the Russian Federation for the IMP UB RAS.

\section*{Conﬂict of Interest}

The authors declare that they have no known competing financial
interests or personal relationships that could have appeared to influence
the work reported in this paper.

\section*{Data Availability Statement}

Data will be made available on request.

\bibliographystyle{elsarticle-num} 
\bibliography{els-ref}

\end{document}